\documentclass[pra,twocolumn,amsmath,amssymb,superscriptaddress]{revtex4-1}
\usepackage{graphicx}

\newcommand{\Toulouse}{Universit\'e de Toulouse ; UPS ; Laboratoire Collisions Agr\'egats R\'eactivit\'e, IRSAMC ; F-31062 Toulouse, France and CNRS ; UMR 5589 ; F-31062 Toulouse, France}
\newcommand{\Liege}{CESAM research unit, University of Li\`ege, 4000 Li\`ege, Belgium}

\usepackage{color}
\usepackage{soul}
\newcommand{\PS}[1]{\textcolor{red}{#1}}
\newcommand{\PSerase}[1]{\textcolor{red}{\st{#1}}}
\newcommand{\PScomment}[1]{\textcolor{red}{\textit{\textbf{((#1))}}}}
\renewcommand{\PS}[1]{#1}
\renewcommand{\PSerase}[1]{}
\renewcommand{\PScomment}[1]{}

\begin{document}

\title{Beyond effective Hamiltonians: micromotion of Bose Einstein condensates in periodically driven optical lattices}

\author{M. Arnal}
\affiliation{\Toulouse}
\author{G. Chatelain}
\affiliation{\Toulouse}
\author{C. Cabrera-Guti\'errez}
\affiliation{\Toulouse}
\author{A. Fortun}
\affiliation{\Toulouse}
\author{E. Michon}
\affiliation{\Toulouse}
\author{J. Billy}
\affiliation{\Toulouse}
\author{P. Schlagheck}
\affiliation{\Liege}
\author{D. Gu\'ery-Odelin}
\affiliation{\Toulouse}

\begin{abstract}
We investigate a Bose Einstein condensate held in a 1D optical lattice whose phase undergoes a fast oscillation using a statistical analysis. The averaged potential experienced by the atoms boils down to a periodic potential having the same spatial period but with a renormalized depth. However, the atomic dynamics also contains a \emph{micromotion} whose main features are revealed by a Kolmorogov-Smirnov analysis of the experimental momentum distributions. We furthermore discuss the impact of the micromotion on a quench process corresponding to a proper sudden change of the driving amplitude which reverses the curvature of the averaged potential.
\end{abstract}

\maketitle

\section{Introduction}

Periodic driving has developed into an indispensable item of the experimental toolbox for the manipulation of ultracold atomic quantum gases \cite{Eck17RMP}. The possibility to subject the magnetic or optical potential that confines the atoms to a time-periodic modulation has opened various perspectives towards the realization of unconventional Hamiltonians, in particular in the context of optical lattices. This has allowed for the development of techniques to dynamically control the superfluid-Mott insulator transition \cite{EckWeiHol05PRL,LigO07PRL}, to engineer Peierls phases and nonabelian gauge fields within optical lattice plaquettes \cite{StrO11S,StrO12PRL,HauO12PRL}, to realize multiple Bose Einstein condensates (BECs)\cite{VorO13PRL}, to generate time crystals \cite{Sac15PRA}, as well as to create synthetic dimensions \cite{PriOzaGol17PRA}, to mention a few examples. A periodic modulation of an optical lattice can, moreover, be employed to investigate the kinetics of phase transitions \cite{ClaFenChe16S,FenO18NP} and to study heating and thermalization effects induced by atom-atom interaction \cite{ReiO17PRL,MicO18NJP}. In the specific case of an optical lattice that is subjected to a relatively rapid periodic phase shaking, with a frequency that is large compared to the inter-well tunneling rate but still small compared to the intra-well oscillation frequency, a renormalization of the hopping matrix element appearing within the single-band description of this lattice can thereby be obtained, which can give rise to coherent destruction of tunneling \cite{GemO05PRL,EckWeiHol05PRL,LigO07PRL,KieO08PRL}.

In most of the above studies, the driving is essentially utilized in order to generate an effective time-independent Hamiltonian which exhibits properties that would be hard, if not impossible, to engineer with a purely static potential configuration. This effective Hamiltonian is technically obtained from a separation of the time scales characterizing the rapidly and slowly evolving degrees of freedom: in the case of a resonant driving, it can be determined through a suitable transformation to a corotating frame. It is then common practice to argue that a time averaging can be performed over the rapidly evolving degrees of freedom which yields the effective time-independent Hamiltonian. 
The above reasoning may be sufficient to qualitatively and quantitatively characterize a large number of experimental configurations and scenarios in the context of periodically driven quantum gases. However, it is incomplete insofar as it does not account for the hardly appreciable but, under certain circumstances, significant phenomenon of \emph{micromotion} \cite{MueO00PRL,AniO15PRB}: the fact that the atoms that are exposed to such a rapid driving undergo a periodic wiggling of tiny amplitude which is synchronized to the external modulation frequency. It must be accounted for in the case of more complicated driving protocols such as quench processes where parameters of the effective Hamiltonian are suddenly varied.

To shed more light on this issue and get a more quantitative grasp of this phenomenon, we consider the most basic driving scenario that one could think of in this context: a rubidium-87 BEC in a one-dimensional (1D) horizontal lattice potential whose phase undergoes a fast oscillation. As detailed in Sec.~\ref{sec1}, time averaging over one driving period gives rise to a renormalization of the lattice amplitude, which thereby can effectively vanish or even reach negative values. This effect can be observed directly on the momentum distribution of the BEC (after a time-of-flight). In Sec.~\ref{sec2}, we show how a more refined statistical analysis of this very same distribution reveals quantitatively the fingerprint of the micromotion. Furthermore, it is known that a sudden change in the fast driving force can also trigger an evolution of the slow degree of freedom \cite{RidWei09PRA,CleHijWal10PRA}. In Sec.~\ref{sec3}, we exemplify this effect by a proper choice of an abrupt change of the amplitude of modulation which reverses the curvature of the averaged potential experienced by the atoms, and we discuss how this phenomenon can be affected by the presence of micromotion.

\section{The averaged potential}
\label{sec1}

The lattice potential reads
\begin{equation}
V(x) = - \frac{s}{2} E_L \cos(k x)
\end{equation}
with $E_L = \hbar^2 k^2/(2 m) = 4 E_{\rm rec}$, where $E_{\rm rec}$ is the recoil energy associated with the photon of the laser that creates the lattice, and $m$ the mass of an atom.  The lattice spacing is $d = 2 \pi / k = 532$~nm and the dimensionless lattice strength is given by $s$.  The horizontal lattice shaking is characterized by the dimensionless amplitude $\varphi_0$ and the frequency $\omega/(2\pi)$. The resulting time-dependent Hamiltonian governing the dynamics of the atoms is given by
\begin{equation}
H(t) = -\frac{\hbar^2 \nabla^2}{2 m} + V\left( x - \frac{2 \varphi_0}{k} \sin(\omega t)\right) + V_0(x,y,z) \,, \label{eq:H}
\end{equation}
where $V_0$ represents an overall harmonic confinement potential with trapping frequencies on the order of $\sim 30$~Hz. We have tested modulation frequencies, $\omega / (2 \pi )$, in the range of $25$ to $500$~kHz. 

For parameters such that $\hbar \omega / (s E_L) \gg 1$, the energy associated with the modulation largely exceeds the barrier height of the lattice wells. In this regime, the time-periodic lattice potential can safely be replaced by its temporal average over one period of the driving. This averaging gives rise to the renormalization of the lattice amplitude in terms of the Bessel function $J_0(\xi) = \pi^{-1}\int_0^\pi \cos(\xi \sin \theta) d\theta$, and the dynamics of the slow degree of freedom is dictated by an effective time-independent Hamiltonian
\begin{equation}
\bar{H} = -\frac{\hbar^2 \nabla^2}{2 m}  - \frac{s_{\rm eff} }{2} E_L \cos(k x) + V_0(x,y,z) \label{eq:Heff} \,.
\end{equation}
 where 
 \begin{equation}
 s_{\rm eff}=s J_0(2\varphi_0) \label{eq:seff}
\end{equation}
is the effective renormalized lattice strength. 
This renormalization phenomenon is easily verified in our experiments. To this end, we load adiabatically the BEC into the phase-modulated lattice by ramping up the lattice strength to its final value $s$ in a few ms. After a 2 ms holding time in
the phase-modulated lattice, we switch off all confinements.  The atomic cloud then freely expands and is imaged after a long time-of-flight ($>$ 25 ms) so to access its momentum distribution. 

\begin{figure}[t]
\begin{center}
\includegraphics[width=\columnwidth]{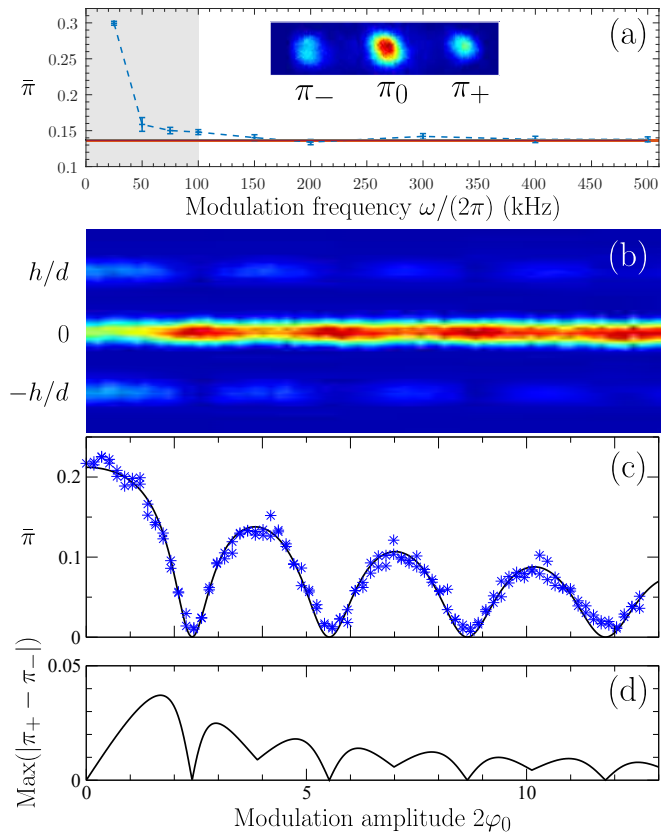}
\end{center}
\caption{\label{fig:1} (Color online). (a) Averaged side-peak population as a function of the modulation frequency for an optical lattice of dimensionless strength $s=7.3\pm 0.1$ and a modulation amplitude $2\varphi_0=3.316$. The horizontal solid line corresponds to the prediction for the static lattice of effective depth $s_{\rm eff}=sJ_0(2\varphi_0)=2.5$ (see text). Blue stars in (c) are the averaged side-peak population of the two side peaks extracted from the time-of-flight images (b) as a function of the modulation amplitude for the fixed modulation frequency $\omega/(2\pi) = 500$ kHz, black solid lines correspond to the theoretical predictions (see text). Panel (d) shows the expected maximal values for the asymmetry of the two side-peak populations. Lattice strength parameter for (b,c,d): $s = 6.4$.}
\end{figure}

The inset of Fig.~\ref{fig:1}(a) shows a typical time-of-flight absorption image. It displays a large peak centered about zero momentum as well as two side peaks centered about the momenta $p = \pm h / d$ \cite{PedO01PRL}. The latter arise from the periodic spatial modulation of the condensate wave function induced by the presence of the lattice potential. In  Fig.~\ref{fig:1}(a), we investigate the validity of the description in terms of an averaged potential for the slow degree of freedom. For this purpose, we plot the averaged population in the side peaks $\bar \pi = (\pi_-+\pi_+)/2$ as a function of the driving frequency $\omega/(2\pi)$ for the parameters $2\varphi_0=3.316$, and $s=7.3\pm 0.1$  \cite{calibration}. We find a good agreement with the prediction  $\bar \pi = 0.136$ associated to a \emph{static} lattice potential of dimensionless strength $s_{\rm eff}=sJ_0(2\varphi_0)=2.5 $ for $\omega > 2\pi \times 100$ kHz while the largest interband transition frequency between the ground state and first excited band is equal to 19.7 kHz for the chosen lattice depth. In the following, we choose $\omega = 2\pi \times 500$ kHz (if not otherwise stated) and $s = 6.4$  \cite{calibration}. 

According to Eq.~\eqref{eq:Heff}, the side peaks shall vanish whenever the driving amplitude $2\varphi_0$ equals a zero of the Bessel function $J_0$. This behaviour is indeed confirmed in Fig.~\ref{fig:1}(b) where we show the measured momentum distribution as a function of the driving amplitude.

More quantitatively, neglecting the effects of atom-atom interactions \footnote{We numerically verified that atom-atom interactions are not relevant in our experiments.} and of the overall harmonic confinement, we can express the condensate wave function associated with the Hamiltonian \eqref{eq:Heff} as
\begin{eqnarray}
\bar{\psi}_0(x ) & = & \sqrt{2} \,\mathrm{ce}_0[k x/2, s_{\rm eff}] \nonumber \\
& = & c_0(s_{\rm eff}) + \sum_{\ell=1}^\infty 2 c_\ell(s_{\rm eff}) \cos(\ell \PS{k x}) \label{eq:psi0}
\end{eqnarray}
(up to some global normalization prefactor), where \PSerase{$\theta=kx$ and }$\mathrm{ce}_0$ is the Mathieu function of the first kind \cite{Mathieu}. The fraction of atoms contained within the central peak and within the two side peaks of the momentum distribution shown in the inset of Fig.~\ref{fig:1}(a) are then given by the Fourier components $\pi_0 = |c_0(s_{\rm eff})|^2$ and $\pi_\pm = |c_1(s_{\rm eff})|^2$, respectively. In Fig.~\ref{fig:1}(c), we plot the averaged side-peak population as a function of the driving amplitude. We find an excellent agreement between the experimental results and the values obtained from the previous analysis.

{\section{Statistical analysis of the atomic micromotion}
\label{sec2}

The above reasoning is incomplete insofar as it does not account for the presence of micomotion. To retrieve the latter, it is useful to derive the effective Hamiltonian \eqref{eq:Heff} in a more rigorous manner in the framework of adiabatic quantum perturbation theory, following the lines of Refs.~\cite{RahGilFis03PRA,GolDal14PRX,LelO17PRX}. In practice, this amounts to subjecting the condensate wave function to a gauge transformation $\psi \mapsto \bar{\psi} = U^\dagger(t) \psi$ with the unitary operator
\begin{equation}
U(t) = \mathcal{T} \exp\left[ - \frac{i}{\hbar} \int_{t_0}^t K(\omega t') dt' \right]
\end{equation}
which is generated by the periodic pseudo-Hamiltonian $K(\theta) = K(\theta + 2 \pi)$ satisfying 
\begin{equation}
\int_{-\pi}^\pi K(\theta) d\theta = 0 \label{eq:K}
\end{equation}
(thus ensuring that $U$ is time-periodic), where $\mathcal{T}$ is the time-ordering operator. The Hamiltonian \eqref{eq:H} of the system is then transformed according to
\begin{eqnarray}
H(t) \mapsto \bar{H}(t) & = & U^\dagger(t) [H(t) - K(t)] U(t) \nonumber \\
& \simeq & H(t) - K(t) + \mathcal{O}(\omega^{-1})
\end{eqnarray}
in the lowest order in the inverse driving frequency. In view of the constraint \eqref{eq:K} we choose 
\begin{equation}
K(\omega t) = H(t) - \bar{H}
\end{equation}
with $\bar{H}$ the time average \eqref{eq:Heff} of the Hamiltonian, and thus obtain $\bar{H}(t) = \bar{H}$ for all $t$.

Micromotion is encoded in the inverse gauge transformation back to the laboratory frame according to $\psi = U(t) \bar{\psi}$. Evaluating
\begin{eqnarray}
\lim_{\theta_0\to-\infty} \int_{\theta_0}^\theta K(\theta') d\theta' & \equiv & \lim_{\epsilon\to 0_+} \int_{-\infty}^\theta K(\theta') e^{\epsilon \theta'} d\theta' \nonumber \\
& = & - \frac{1}{2\pi} \int_{-\pi}^{\pi} \theta' K(\theta - \theta' - \pi) d\theta
' \end{eqnarray}
by means of Eq.~\eqref{eq:K}, we obtain
\begin{eqnarray}
U(t) & \simeq & 1 - \frac{i}{\hbar \omega} \int_{\omega t_0}^{\omega t} K(\theta') d\theta' \nonumber \\
& = & 1 + \frac{i}{2\pi\hbar\omega} \int_{-\pi}^{\pi} \theta' V[ x + \frac{2\varphi_0}{k} \sin(\omega t - \theta')] d\theta'
\end{eqnarray}
in the lowest nontrivial order in $\omega^{-1}$.
Using the Fourier series expansion $\exp(i \xi \sin \theta) = \sum_{n=-\infty}^\infty J_n(\xi) \exp(i n \theta)$ with $J_n(\xi) = (2\pi)^{-1} \int_{-\pi}^\pi \exp(i \xi \sin \varphi - n \varphi) d\varphi$ the Bessel function of the first kind of order $n$, we obtain the wave function in the laboratory frame as
\begin{eqnarray}
&\psi & (x,t)  =  \left\{1 + \frac{s \hbar k^2}{8 m \omega} \sum_{n=1}^\infty \left[ e^{i n \omega t} - (-1)^n e^{- i n \omega t} \right] \right.\nonumber\\
&& \times \left. \frac{J_n(2 \varphi_0)}{n} \left[ (-1)^n e^{i k x} + e^{- i k x} \right] \right\} \bar{\psi}_0(x)   \label{eq:micromotion}
\end{eqnarray}

Clearly, the presence of micromotion introduces additional contributions of oscillatory nature to the side-peak populations, which are then evaluated from the spatial Fourier series of Eq.~\eqref{eq:micromotion} as
\begin{eqnarray}
&\pi_\pm&(t)  =  \left| c_1(s_{\rm eff}) + \frac{s \hbar k^2}{8 m \omega} \sum_{n=1}^\infty \left[ e^{i n \omega t} - (-1)^n e^{- i n \omega t} \right] \right. \nonumber\\ 
&&\times \left. \frac{J_n(2 \varphi_0)}{n} (\mp 1)^n \left[ c_0(s_{\rm eff}) + (-1)^n c_2(s_{\rm eff}) \right] \right|^2.
\end{eqnarray}
These additional contributions with respect to $|c_1(s_{\rm eff})|^2$ are negligibly small on temporal average as well as for the mean side-peak population $(\pi_+ + \pi_-)/2$, where they scale quadratically with $s \hbar k^2/(8 m \omega) \sim 0.026$. However, they give rise to significant oscillatory side-peak population imbalances 
\begin{eqnarray}
\pi_+(t) - \pi_-(t) & = & - c_1(s_{\rm eff}) [ c_0(s_{\rm eff}) - c_2(s_{\rm eff}) ] \frac{s \hbar k^2}{m \omega} \nonumber \\ & \times & \sum_{\ell=0}^\infty \frac{J_{2\ell+1}(2 \varphi_0)}{2\ell+1} \cos[(2\ell+1)\omega t] \label{eq:pdiff}
\end{eqnarray}
whose maximal absolute values within one driving period are displayed in Fig.~\ref{fig:1}(d) as a function of the driving amplitude.

\begin{figure}[t]
\begin{center}
\includegraphics[width=\columnwidth]{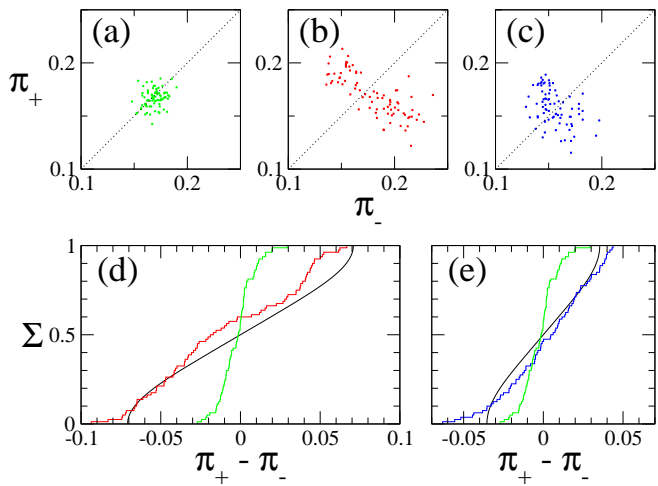}
\end{center}
\caption{\label{fig:2}
(Color online) Upper row: measured populations $\pi_\pm$ of the two side peaks (a) in the absence of the modulation ($s=3.1$, $\bar\pi=0.156$) and, in the presence of modulation with a frequency of $\omega/(2\pi) = 250$ kHz (b) and $500$ kHz (c) and for the lattice strength $s=6.6$ i.e. $s_{\rm eff}=s/2=3.3$ ($\bar\pi=0.162$). For these two latter cases, the modulation amplitude $2 \varphi_0 = 1.520$ was chosen such that a significant amplitude of micromotion would be obtained [see Fig.~\ref{fig:1}(d)]. The corresponding cumulative distribution functions extracted from the experimental data for $\omega/(2\pi)  = 250$ kHz and $500$ kHz are shown in panels (d) and (e) (blue and red staircase-like lines), respectively. While an error function profile would be expected in the absence of the driving (as shown by the light green distribution obtained for $\varphi_0 = 0$), the actual distribution functions agree very well with their analytical predictions based on the theory of micromotion (smooth black lines).}
\end{figure}

In our experiment, it is not possible to directly verify Eq.~\eqref{eq:pdiff}. Indeed, the switch-off time of the lattice and confinement potentials is not synchronized with the driving period. However, we can perform a statistical analysis of the experimentally detected side-peak population imbalances in the spirit of a Kolmogorov-Smirnov test, assuming a uniform probability distribution for the final phase $\omega t$ at the instant when all confinements are switched off. The results of such a statistical analysis are shown in Fig.~\ref{fig:2} for the driving amplitude $2\varphi_0 = 1.520$ that leads to the largest side-peak population imbalance according to Fig.~\ref{fig:1}(d) and for the driving frequencies $\omega/(2\pi) = 500$~kHz and $250$~kHz. Very good agreement is found between the experimental cumulative probability distribution, $\Sigma$,  to detect a given side-peak population imbalance and its analytical prediction which is essentially obtained from an inversion of Eq.~\eqref{eq:pdiff} within $0 \leq \omega t \leq \pi$. In the absence of modulation (see Fig.~\ref{fig:2}(a)), the imbalance distribution is Gaussian and its cumulative probability is therefore an error function. In the presence of the modulation, the $\Sigma$ prediction resembles an arccosine profile given the fact that the terms with $\ell>0$ in the sum in Eq.~\eqref{eq:pdiff} are negligibly small for the driving amplitude under consideration. The width of the $\Sigma$ function is expected to vary as $\omega^{-1}$, a property that can be directly seen in Figs.~\ref{fig:2}.

\section{Engineering the micromotion}
\label{sec3}

\begin{figure}[t]
\begin{center}
\includegraphics[width=\columnwidth]{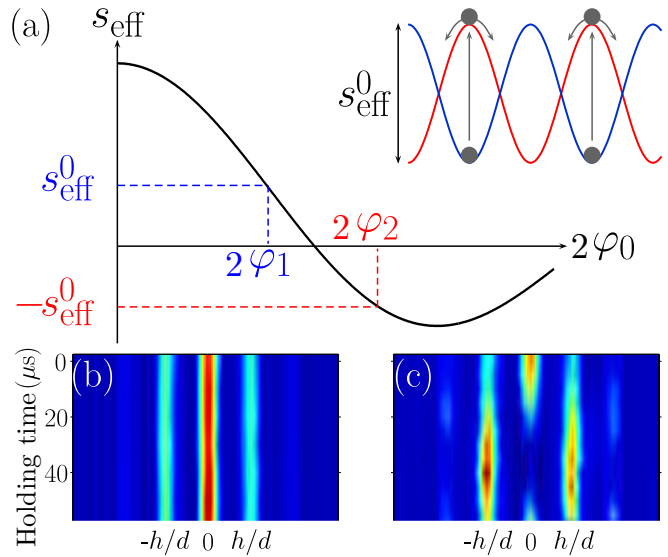}
\end{center}
\caption{\label{fig:3}
(Color online). (a) Effective lattice strength as a function of the modulation amplitude $\varphi_0$, and sketch of the experiment in which the minima of the periodic potential are suddenly replaced by maxima as a result of a modulation amplitude change from $\varphi_1$ to $\varphi_2$. (b-c) Time-evolution of the interference pattern obtained after time-of-flight: (b) for a BEC loaded and remaining in a phase-modulated lattice with $2\varphi_1=1.920$ (c) after the sudden change of the modulation amplitude from $2\varphi_1$ to $2\varphi_2=3.142$. Both amplitudes of modulation correspond to close effective lattice depth. The change from  $\varphi_1$ to $\varphi_2$ triggers an out-of-equilibrium dynamics since the wave function peaks initially at the bottom of lattice wells are abruptly placed at the top of the periodic potential hills where they split into two packets with opposite momenta (c).}
\end{figure}

Although clearly visible in our experiments, the effect of micromotion can still be considered to have a perturbative impact for the averaged dynamics of the atoms within the periodically modulated lattice potential under consideration. However, it must be accounted for in the presence of nontrivial quench processes where parameters of the effective Hamiltonian \eqref{eq:Heff} are suddenly varied. Such a quench process can, e.g., consist in an inversion of the effective lattice strength \eqref{eq:seff} which can be realized through a sudden steering of the modulation amplitude $2\varphi_0$ across a zero of the Bessel function $J_0$. 


The possibility of reversing the curvature of the averaged potential by engineering the micromotion is not a general property of rapidly varying potentials. For instance, it does not occur for a harmonic potential whose position would be modulated \cite{RidWei09PRA}.
In our case, it appears mathematically as a consequence of the renormalization factor that involves a zeroth order Bessel function whose sign depends upon its argument. Once suddenly placed at the top of the potential hills, one expect the packet to split into two symmetric wave packets that move into opposite directions. 
It is worth noticing that such a sudden change of the depth and phase of the lattice can be done without modifying the light intensity of the beams that generate the lattice. This quench can therefore be performed in the presence of an active intensity locked system.

The corresponding experimental results are reported in Fig.~\ref{fig:3} where we show the measured momentum distribution as a function of the holding time within the inverted lattice after the quench. The sudden change of the modulation amplitude has been calculated to realize  a change of the sign of the lattice amplitude while keeping the lattice strength to nearly the same value. We clearly observe the splitting revealing that the sign of the averaged potential experienced by the atoms has been indeed changed. 

A more careful look reveals the presence of a slight asymmetry in the split packets as they evolve in time. This asymmetry could reveal an initial small oscillation of the condensates in each well of the optical lattice or a small shift of the lattice when the phase shift is performed. It is worth noticing that such an asymmetry can also arise in a perfectly symmetric configuration, namely due to the presence of micromotion. Indeed, if the quench takes place at an instant $t$ where the side-peak population imbalance $\pi_+(t) - \pi_-(t)$ is non-zero, the wavefunctions shall split asymmetrically. As we do not control the quench time at the scale of the micromotion period, each measurement samples the distribution of the population imbalance, an effect that results for our parameter to a slight asymmetry but with an amplitude smaller than the one observed in the experiment.

\section{Conclusion}

In summary, we have shown how a phase-modulated optical lattice can serve as a testbed for verifying the presence of micromotion. The latter can be directly seen in the momentum distribution of the atoms that formed a Bose Einstein condensate within the modulated lattice, by performing a statistical comparison of the population difference of the positive- and negative-momentum side peaks with the analytical prediction resulting from the theory of micromotion. We have explained how a sudden change of the fast driving phase can have a strong impact on the slow atomic motion. We have furthermore discussed how the micromotion could contribute to the symmetry breaking between wave packets that emerge from an inversion of the effective lattice amplitude. A more refined verification of such an effect will require a more precise control of the quench time on the scale of the driving period. This will then open various perspectives for utilizing the concept of micromotion as an engineering tool for the control of ultracold quantum gases.}

\begin{acknowledgements}
We acknowledge support from Programme Investissements d'Avenir under the program ANR-11-IDEX-0002-02, reference ANR-10-LABX-0037-NEXT and its invited researcher program, and the research funding grant ANR-17-CE30-0024. M.~A.\ acknowledges support from the DGA (Direction G\'en\'erale de l'Armement). 
\end{acknowledgements}

\bibliography{micromotion}

\end{document}